\documentclass{article}
\usepackage{ltwol2e}

\def\Journal#1#2#3#4{{#1} {\bf #2}, #3 (#4)}

\def\NPB{{\em Nucl. Phys.} B}
\def\NPP{\em Nucl. Phys. Proc. Suppl.}
\def\PLB{{\em Phys. Lett.}  B}
\def\PRD{{\em Phys. Rev.} D}
\def\PRL{\em Phys. Rev. Lett.}
\def\RMP{\em Rev. Mod. Phys.}
\def\CMP{\em Comm. Math. Phys.}

\def\JPQ{\em J. Physique}
\def\SMF{\em Bull. Soc. Math. France}

\def\be{\begin{equation}}
\def\ee{\end{equation}}
\def\bea{\begin{eqnarray}}
\def\eea{\end{eqnarray}}

\def\pad{\partial}
\def\pas{\partial\!\!\!/}
\def\Dsl{D\!\!\!\!/\,}

\def\ovr{\over}

\def\til{\tilde}
\def\pri{^\prime}

\def\be{\beta}
\def\ga{\gamma}
\def\de{\delta}
\def\ep{\epsilon}

\def\et{\eta}
\def\th{\theta}

\def\la{\lambda}
\def\rh{\rho}

\def\si{\sigma}

\def\ps{\psi}

\def\psb{\overline{\psi}}
\def\etb{\overline{\eta}}

\def\beq{\begin{equation}}
\def\eeq{\end{equation}}
\def\bdm{\begin{displaymath}}
\def\edm{\end{displaymath}}
\def\bea{\begin{eqnarray}}
\def\eea{\end{eqnarray}}

\begin{document}

\title{THEORETICAL ASPECTS OF TOPOLOGICALLY UNQUENCHED QCD}

\author{S. D\"{u}rr}

\address{University of Washington, Physics Department, Box 351560, 
Seattle, WA 98195, U.S.A.\\E-mail: durr@phys.washington.edu}   


\twocolumn[\maketitle\abstracts{
I give an outline of my recent proposal to take the QCD functional determinant
in lattice simulations partially into account: The determinant is split into
two factors, the factor referring to a standard background in each topological
sector is kept exactly, the factor describing the effect of the smooth deviation
of the actual configuration from the reference background is replaced by one.
The issue of how to choose the reference configurations is discussed and it is
argued that ``topologically unquenched QCD'' is an interesting starting point
to study full QCD in lattice simulations as it gets the main qualitative
features right from the beginning.
}]


\section{What's wrong with quenched QCD ?}


In QCD the fermion functional determinant is a nonlocal contribution to the
gluon effective action. This nonlocality has an unpleasant effect in lattice
calculations as it slows down present numerical algorithms dramatically when
the quark-masses get small.

In order to elude this problem most numerical simulations in the past have been
done in the ``quenched approximation'' where the determinant is simply replaced
by one \cite{Hamber}.
More recently, simulations in the ``partially quenched approximation'' have
become available \cite{SESAM}, where the determinant gets
evaluated at quark masses which are higher than those in the propagators.
Thus (partial) quenching amounts to suppressing the contribution of all
internal fermion loops in QCD by giving the quarks unphenomenologically high
or infinite masses.

Attempts to introduce the corresponding modifications in the low-energy
theory artificially in order to learn how to correct for them --~the results
being ``quenched'' and ``partially quenched Chiral Perturbation Theory''~--
have shown that the (partially) quenched approximation is, in some aspects,
fundamentally different from the full theory:
First of all, numerical results won in quenched or partially quenched
simulations should (at least in principle) be corrected for the occurrence of
``enhanced chiral logarithms'' \cite{QuenchedQCD,PartiallyQuenchedQCD}.
In addition, the $\et\pri$ was found to be a pseudo-Goldstone boson in
quenched QCD (as opposed to the situation in QCD, where it is heavier than the
lowest-lying octet of pseudo-scalar pseudo-Goldstone mesons by
more than a factor $\sqrt{3\,}$) and its propagator shows --~in case the
low-energy analysis is correct~-- a pole of order two (which spoils any
field-theoretic interpretation) right at the same position in the $p^2$-plane
where its first-order pole is \cite{QuenchedQCD,PartiallyQuenchedQCD}.
Even without particle-interpretation, quenched QCD is strictly confined to
Euclidean space-time; there is no continuation of its Green's functions to
Minkowski space-time \cite{Morel}.

Thus it seems legitimate to search for an alternative starting point for
approaching full QCD which gets some qualitative aspects of
full QCD right from the beginning.


\section{What is ``topologically unquenched QCD'' ?}


The aim is to identify a part of the functional determinant which is cheap from
the computational point of view but essential from the field-theoretical point
of view.

We start from the generating functional of (eucli\-dean) QCD in the form
where the fermionic degrees of freedom have been integrated out
\bea
Z_\th^{\rm QCD}[\etb,\et]&=&
N\cdot\int\!DA\;{\det(\Dsl\!+\!M)\ovr\det(\,\pas\,\!+\!M)}
\nonumber
\\
&{}&
e^{\etb(\Dsl+M)^{-1}\et}\;
e^{-\!\int\!{1\ovr4}GG\;+i\th\!\int\!{g^2\ovr32\pi^2}G\til G}
\label{1}
\eea
where $\Dsl=\ga_\mu(\pad_\mu-igA_\mu)$ is the (euclidean) Dirac operator and
$\til G_{\mu\nu}={1\ovr2}\ep_{\mu\nu\si\rh} G_{\si\rh}$ the dual of the
field-strength operator and where the measure $DA$ includes gauge-fixing and
Faddeev Popov terms. In (\ref{1}) a factor which does not depend on the
gauge field to be integrated over has been pulled out of the normalizing
factor and the convention is that the quark mass matrix $M$ is diagonal
and of rank $N_{\!f}$ with the CP-violation stemming entirely from $\th$
(if $\th\neq\pi{\bf Z}$); the shorthand-notation to be used here and in
subsequent formulas is $\det(\!\Dsl\!+\!M)=\prod_{i=1}^{N_f}
\det(\!\Dsl\!+\!m_i)$ in the determinant and $\etb(\Dsl\!+\!M)^{-1}\et=
\sum_{i=1}^{N_f}\etb_{(i)}(\Dsl\!+\!m_i)^{-1}\et_{(i)}$ in the propagator.

QCD is known to show a topological structure \cite{Topology}: The
$SU(3)$-gauge-field configurations on ${\bf R}^4$ with finite action boundary
condition or on the torus ${\bf T}^4$ fall into inequivalent topological
classes (labeled by an index $\nu=g^2/32\pi^2\cdot\int G\til G\,dx \in{\bf Z}$).
In a given sector $\nu$, any two configurations may be continuously deformed
into each other but not into any configuration with a different index $\nu$.
Due to this topological structure, the integral in (\ref{1}) may be rewritten
as a sum of integrals over the individual sectors
$\int\!DA\to\sum_{\nu\in{\bf Z}}\int\!DA^{(\nu)}$
and the determinant factorizes ($\Dsl=\Dsl^{(\nu)}$)
\beq
{\det(\Dsl\!+\!M)\ovr\det(\,\pas\,\!+\!M)}=
{\det(\Dsl^{(\nu)}_{\rm std}\!+\!M)\ovr\det(\,\pas\,+M)}\cdot
{\det(\Dsl^{(\nu)}\!+\!M)\ovr\det(\Dsl^{(\nu)}_{\rm std}\!+\!M)}
\label{2}
\eeq
where the first factor depends on $\nu$ only and thus may be pulled out of the
integral.
In this form it is obvious that quenched QCD actually does two modifications:
It sets both determinant factors equal one, whereas partially quenched QCD
keeps both of them at the price of using unrealistically high quark masses.

In ``topologically unquenched QCD'' the first factor in (\ref{2}) is kept
exact (with the same quark mass as in the propagator between $\etb,\et$) and
only the second factor is replaced by one \cite{TUQCD}, i.e. the ``theory''
is defined through
\bea
Z_{\th,\{A^{(\nu)}_{\rm std}\}}^{\rm TU-QCD}[\etb,\et]
&=&
N\cdot\sum_{\nu\in{\rm\bf Z}}
{\det(\Dsl^{(\nu)}_{\rm std}\!+\!M)\ovr\det(\Dsl^{(0)}_{\rm std}\!+\!M)}\
e^{i\nu\th}\cdot
\nonumber
\\
&{}&
\int\!\!D\!A^{(\nu)}\;e^{\etb(\Dsl^{(\nu)}\!+M)^{-1}\et}\;
e^{-\!\!\int\!{1\ovr4}GG}\;.
\label{3}
\eea

At this point, the motivation to treat the two determinant factors in (\ref{2})
on unequal footing is just an economic one: The ``topological'' factor is
universal for all configurations within one class and bears the knowledge
about the nontrivial topological structure of QCD.
On the other hand, the ``continuous'' factor in (\ref{2}) causes a dramatic
slowdown in numerical simulations as this part of the determinant (or its
change) has to be computed for each configuration individually.

Note that, in contrast to the full-QCD generating functional (\ref{1}),
the ``topologically unquenched'' truncation (\ref{3}) does depend on the
choice of reference-backgrounds.
Later, two alternative strategies of how to choose, in a given sector,
the reference-configuration are described and such a choice of strategy,
once it is done, is considered part of the definition of the theory.
Obviously there is no a-priori evidence that --~with any of these two
choices~-- the approximation (\ref{3}) should be particularly good.
Nevertheless, known analytical knowledge about QCD in a finite box seems to
indicate that including the ``topological'' part of the determinant is
sufficient to get some basic features of full QCD qualitatively right.


\section{How to implement ``topologically unquenched QCD'' ?}


Obviously, for ``topologically unquenched QCD'' to be practically useful, at
least two requirements have to be fulfilled:
First, a recipe of how to make a good choice for the reference backgrounds
in (\ref{3}) is needed.
Second, the costs in terms of CPU-time for separating the ``topological'' part
of the determinant from the remainder have to be smaller than the costs would
be to compute the determinant as a whole.
These issues shall be discussed.


\subsection{Upgrading a quenched sample}

A quenched sample may get modified to be representative in the
sense of ``topologically unquenched QCD'':
\begin{enumerate}
\item
Use a method you consider both trustworthy and efficient to compute for each
configuration its topological index $\nu$.
\item
Use the gauge action you trust to compute, in each class, the gauge-action of
every configuration as well as the class-average $\bar S^{(\nu)}$ and choose the
reference-configuration according to one of the following two prescriptions:
\begin{itemize}
\item[($i$)]
Choose --~out of the class $\nu$~-- the configuration with minimal gauge-action
as the representative $A^{(\nu)}_{\rm std}$.
\item[($ii$)]
Choose --~out of the class $\nu$~-- the configuration for which the
gauge-action is closest to the class-average $\bar S^{(\nu)}$ as the
representative $A^{(\nu)}_{\rm std}$.
\end{itemize}
\item
Use the fermion action and the method you consider both trustworthy and
efficient to compute the determinants
$\det((\Dsl^{(\nu)}_{\rm std}\!+\!M)/((\Dsl^{(0)}_{\rm std}+\!M))$.
\item
For any higher topological sector either include the corresponding
determinant computed in step 3 into the measurement or eliminate the
corresponding fraction of configurations from that sector.
\end{enumerate}
A few points need immediate clarification:

The first problem is that on the lattice, the topological structure of the
continuum-theory is washed out; simply computing $\nu=g^2/32\pi^2\cdot\int
G\til G\,dx$ gives a value for $\nu$ which is, in general, not an integer.
A solution to this problem is so crucial to the overall performance of the
``topologically unquenched'' approximation that it shall be discussed in
a separate subsection.

The second point is that for computing the determinant ratio in step 3
one cannot rely on any method which is tantamount to an expansion in $\de A$,
since the two backgrounds are far from each other.
The necessary ab-initio computation is achieved e.g. by the eigenvalue
method: typically the first few hundred eigenvalues of the Dirac operator
on a given background may be determined.

Finally: What's the difference in terms of physics between the two strategies
of how to choose the reference-backgrounds in step 2 ?
We emphasize that for either choice there is a sound theoretical motivation.
Strategy ($i$) --~choose the configuration which minimizes the gauge action~--
is nothing but the semiclassical ansatz being pushed to account for topology:
Within each sector, the determinant is exact for the configuration having
least gauge-action, i.e. for the one which, in a semiclassical treatment,
gives the dominant contribution of that sector to the path-integral.
Strategy ($ii$) --~choose the configuration which, in its gauge-action, is
closest to the class-average of that sector~-- takes into account that
the Monte Carlo simulation as a whole doesn't try to minimize the total
action density but rather the free-energy density:
The configuration which is most typical in a certain sector is not the one
with minimal action but the one which has additional instanton-antiinstanton
pairs plus topologically trivial excitations such as to find an optimum between
the additional amount of action to be paid and the additional amount of entropy
to be gained.
It is the very aim of the second strategy to choose in each sector a
``most-typical'' background (which realizes such an optimum pay-off) as
reference-configuration.
It should be stressed that even though the two strategies end up
selecting reference-backgrounds which look highly different (exceedingly
smooth in the first case versus pretty rough in the second case) the final
results may still be close to each other -- the only thing which matters
is the (strategy-intrinsic) determinant ratio computed in step~3.


\subsection{Generating a ``topologically unquenched'' sample}

The fact that the ``topological'' determinant in (\ref{3}) is a number which
depends only on the total topological charge of the actual configuration
but not on its other details suggests that one could try to precompute these
``topological'' determinants on artificially constructed backgrounds prior to
running the simulation.

Within the strictly semiclassical strategy which is reflected by choice~($i$)
the reference-backgrounds are gotten in a rather simple way:
Place $\nu$ copies (for $\nu\!>\!0$) of a single-instanton solution with
typical radius (i.e. $\rh\!\simeq\!0.3{\rm fm}$, cf.~\cite{SchaferShuryak}) on
the lattice and minimize the action with respect to variation of their relative
orientations and positions.
Within the more realistic strategy of choice~($ii$) which accounts for the
competing effects of increased action versus increased entropy the
reference-backgrounds are constructed as follows:
Place $\nu$ instantons (for $\nu\!>\!0$) with typical sizes (i.e.
$\rh\!\simeq\!0.3{\rm fm}$, cf.~\cite{SchaferShuryak}) randomly on the lattice
plus additional instanton-antiinstanton pairs such as to achieve a total
instanton density of $1{\rm fm}^{-4}$ (cf.~\cite{SchaferShuryak}).
Optionally, this background may be dressed with thermal fluctuations by
applying a reasonable number of heating-steps (monitoring $\nu$ in order to
guarantee that it stays unchanged).

Having defined the standard backgrounds in this way a pure Metropolis algorithm
generating a ``topologically unquenched'' sample employs the following steps:
\begin{enumerate}
\item
Make use of an updating procedure to propose a new configuration and
determine its topological index (via the method you trust).
\item
If the configuration realizes a previously unseen $\nu$: Evaluate the
functional determinant on the standard background constructed for that $\nu$
according to your choice of strategy.
\item
Base the decision on whether to accept the proposed configuration on
\beq
\Delta S=
S^{(\nu)}_{\rm new}-
S^{(\nu)}_{\rm old}-
\log(
{\det(\Dsl^{(\nu,{\rm new})}_{\rm std}+M)\ovr
\det(\Dsl^{(\nu,\,\,{\rm old}\,)}_{\rm std}+M)}
)
\label{4}
\eeq
where $S^{(\nu)}_{\rm new/old}$ denotes the gluonic action of the newly
proposed / last accepted configuration.
\end{enumerate}
There is one conceptual deficit this algorithm suffers from:
The procedure for constructing the standard-backgrounds makes use of knowledge
about the size-distribution and partly about the density of (anti-) instantons
which was won in previous lattice-studies.
In other words: The ``topologically unquenched'' simulation as outlined above
is not entirely from first principles.
Moreover, the quality of the final sample depends on how appropriate the
artificial backgrounds are which were used in the computation of the
standard determinants.
Constructing these reference backgrounds is particularly demanding within
strategy ($ii$) as it means that one has to do an a-priori guess which
configuration, within a given sector, is ``most-typical'' in the sense
of full QCD.
Even if ``most typical'' is translated into a technical criterion (e.g.
the configuration which, in its total effective action is closest to the
corresponding class-average of a finite full-QCD sample) there is no
other algorithmic solution to this problem than by doing a full-QCD simulation.
Thus it is reasonable to construct the reference backgrounds as indicated
above, thereby making use of existing knowledge concerning size-distribution
and density of instantons in QCD -- in particular as there is, for strategy
($ii$), a final check concerning the quality of the artificial backgrounds:
In case the guess would have been just perfect, the a-priori ratios of
determinants evaluated on these backgrounds (i.e. the ratios which got used
in the ``topologically unquenched'' simulation) would perfectly agree with the
analogous a-posteriori determinant ratios evaluated on the ``most-typical''
configurations as produced by the run.
Accordingly, a procedure one might think of trying in case the agreement turns
out to be less than satisfactory is just to start over with the simulation --
but this time using the ``most-typical'' configurations found in the first
run rather than the artificial guesses.


\subsection{Measuring topological indies}

Determining the topological index of a newly proposed background in a way which
is fast and reliable is so crucial to the overall-performance of
``topologically unquenched QCD'' as to justify few remarks about this point.

There are several methods \cite{LatticeIndex} to determine, for a given
configuration, its topological index $\nu$.
Some of them were recently compared and found to give -- when implemented with
sufficient care -- to comparable results for the topological susceptibility
\cite{IndexComparison}.
Nevertheless, it is clear that in the present case, where nothing is known
about the spectrum of the Dirac operator on the background at hand, the
field-theoretic method is likely to determine $\nu$ in the quickest possible
way. However, the fact that on the lattice the relevant operator
undergoes thermal renormalization provides a challenge:
Simply integrating the Chern density, i.e. computing $g^2/(32\pi^2)\int
G^a_{\mu\nu}\til G^a_{\mu\nu}\;dx$ gives a value which is, in general, not
close to an integer; in fact, a histogram-plot over many configurations tends
to reveal accumulations near regularly displaced, non-integer values, e.g. near
$0,\pm0.7,\pm1.4$ etc.
There are two options of how to deal with this situation:

The first, simplistic, approach is just to define a ``confidence interval''
--~e.g. $\pm0.2$~-- around each of the values $0,\pm0.7,\pm1.4,\ldots$ and to
assign the configurations lying within these bounds the indices $\nu=
0,\pm1,\pm2\ldots$ etc. The remaining configurations which didn't get an index
assigned are then simply tossed away.

The second, more sophisticated, approach is to make use of the fact that
cooling a configuration is able to remove the effect brought in by thermal
renormalization: Cooling a set of gluon-configurations results in the peaks
(in the histogram plot) being shifted closer to the corresponding integers
and the valleys between the peaks getting thinned out under each sweep.

The problem is that these two methods do not necessarily agree in their results
for a given configuration -- a fact which can be understood on rather simple
grounds:
Under repeated cooling with the naive (Wilson) action, a single-instanton
solution shrinks monotonically until it finally falls through the grid.
In order to prevent the cooling algorithm at least from loosing the large
instantons one has to modify the action w.r.t. which cooling is done in such
a way that all instantons with a radius $\rh$ above a certain $\rh_{\rm thr}$
tend to get blown up (``over-improved action'') or stay constant (``perfect
action'') under a sweep, where typically $\rh_{\rm thr}\simeq 2.3a$.
From the evidence given in \cite{Forcrand} how quickly cooling with an
``over-improved'' action tends to pin down $g^2/(32\pi^2)\int G^a_{\mu\nu}\til
G^a_{\mu\nu}\;dx$ near an integer (say 5 sweeps to be within 2.99 and 3.01,
etc.) performing $O(3)$ ``over-improved'' cooling sweeps seems to be sufficient
to get an unambiguous assignment.
The price to pay, however, is that the small instantons ($\rh<\rh_{\rm thr}$)
get compressed and finally pushed through the grid even more efficiently than
under cooling with an unimproved action \cite{Forcrand}.
From these consideration we conclude that the field-theoretic method with
cooling yields, once it has stabilized, a correct assignment for the
latter cooled configuration which, however, isn't necessarily appropriate for
the initial configuration which may have contained small ($\rh<2.3a$)
instantons.
On the other hand, determining the topological index by the first (simplistic)
approach (no cooling being involved) has an inferior signal-to-noise ratio
(about half of the configurations can't get assigned an index and have to be
tossed away) but for the remaining ones the procedure is sensitive
to all instantons the lattice can support (i.e.~$\rh>0.7a$).


\subsection{Rudimentary cost analysis}

The overhead as compared to a quenched simulation results from the CPU-time
spent on determining $\nu$ for every newly proposed configuration and from the
determinants which get evaluated.
Preparing the reference-backgrounds and computing the determinants is a fixed
investment which is given by $L, a, m$ only (i.e. independent of the length
of the simulation) and evaluating the $O(10)$ standard determinants (for
nowadays typical values of $m$ and $L$) is pretty cheap~\cite{TUQCD}.
On the other hand, determining for each configuration its index $\nu$ gives
rise to costs which grow linearly in simulation-time and thus provide the main
overhead (as compared to Q-QCD) in a long run.
As a consequence, the method for determining the topological
index will have the greatest impact on the overall-performance in TU-QCD.
We have advocated choosing a field-theoretical definition with little or no
cooling at all, which means that either $O(3)$ cooling-sweeps are performed
or 50\% of the configurations have to be tossed away.
Accordingly, in an approximation where a cooling-sweep is considered twice as
expensive as a complete update, the overhead from $\nu$-determinations is
roughly a factor 2...6 over a quenched simulation.
Thus doing a ``topologically unquenched'' run might be considered an
alternative to a high-statistics quenched run.


\section{What about qualitative features of ``topologically unquenched QCD'' ?}


In ``topologically unquenched QCD'' a determinant is introduced which --~as
is seen from (\ref{4})~-- only influences the relative weight of the
different topological sectors; within each sector there is no difference
to quenched QCD.

For QCD in a finite box Leutwyler and Smilga have shown that in the regime
\footnote{$\Sigma\!=\!\lim_{m\rightarrow0}\lim_{V\rightarrow\infty}
|\langle\psb\ps\rangle|$, where $m_i\!=\!m\;\forall i$ for simplicity;
note that $V\Sigma m\rightarrow\infty$ when $m\rightarrow 0$ as the box has to
be scaled accordingly: $L\simeq 1/M_\pi, M_\pi^2\simeq \Lambda_{\rm had}m$.}
$V\Sigma m\gg1$ the distribution of topological indices is gaussian with width
\cite{LeutwylerSmilga}
\beq
\langle\nu^2\rangle=V\Sigma m/N_f
\quad.
\label{5}
\eeq
In quenched QCD the corresponding distribution is much broader as there is
no determinant which suppresses the higher sectors.
In ``topologically unquenched QCD'' the standard determinants result in the
higher sectors being suppressed as compared to a quenched sample but the amount
of suppression strongly depends on the strategy for selecting or constructing
the reference backgrounds.

In strategy ($i$) a sectorial determinant is introduced which is exact
for the background which --~from the classical point of view~-- dominates that
sector. The point is that this semiclassical treatment is indeed justified for
sufficiently small coupling-constant, i.e. in a ridiculously small box where
the topological distribution in QCD is known to be extremely narrow
\cite{LeutwylerSmilga}.
As the box-volume increases the effective coupling gets stronger and strategy
($i$) is unable to account for this change.
To see this more clearly we stipulate the validity of the index theorem on the
lattice \cite{LatticeIndex} which allows us to rewrite the two factors
in~(\ref{2}) using the Vafa-Witten representation \cite{VafaWitten}
\bea
{\det(\Dsl^{(\nu)}_{\rm std}\!+\!M)\ovr\det(\Dsl^{(0)}_{\rm std}\!+\!M)}&=&
\prod_{i=1}^{N_{\!f}}\;m_i^{|\nu|}\cdot
{\prod_{\la>0}\;(\la^{(\nu)\,2}_{\rm std}+m_i^2)\ovr
\prod_{\la>0}\;(\la^{(0)\,2}_{\rm std}+m_i^2)}
\label{6}
\\
\nonumber
\\
{\det(\Dsl^{(\nu)}\!+\!M)\ovr\det(\Dsl^{(\nu)}_{\rm std}\!+\!M)}&=&
\prod_{i=1}^{N_{\!f}}\;
{\prod_{\la>0}\;(\la^{(\nu)\,2}+m_i^2)\ovr
\prod_{\la>0}\;(\la^{(\nu)\,2}_{\rm std}+m_i^2)}
\qquad.
\label{7}
\eea
Strategy ($i$) retains a determinant which is appropriate in a small volume
and thus strongly suppresses the higher topological sectors.
As it comes to larger volumes, the semiclassical treatment breaks down and
the quantum fluctuations packed into the ``continuous'' determinant (\ref{7})
prove able to milder the suppression -- in full QCD, but not within strategy
($i$).
The virtue of strategy ($ii$) is that this change is accounted for
by successively redefining the standard-backgrounds used in (\ref{6}).
In other words: Within strategy ($ii$) parts which would belong to
(\ref{7}) in ($i$) are gradually reshuffled into the
``topological'' part (\ref{6}) as the box-volume increases.
As a consequence, either strategy is supposed to be trustworthy as long as
$V\Sigma m\leq 1$, but only strategy ($ii$) may give a reasonable approximation
to full QCD in the regime $V\Sigma m\gg1$.

Comparing the two factors (\ref{6}) and (\ref{7}) one ends up realizing that
the ``topological'' determinant (\ref{6}) has exactly the same structure as its
QCD counterpart (the latter comes without the subscript ``std'' in the
numerator): The essential ingredient is the prefactor $m^{|\nu|}$.
In QCD, this prefactor is known \cite{LeutwylerSmilga} to cause the strong
suppression of nonzero indices in the limit $V\Sigma m\ll 1$.
The fact that it is still around in the ``topologically unquenched''
approximation (with either choice for the reference-backgrounds) means that
TU-QCD (unlike Q-QCD) shows the phenomenon of chiral symmetry restoration
if the chiral limit is performed in a finite box.

Finally, the fact that the number of virtual quark-loops is not restricted
in TU-QCD means that there is an infinite number of diagrams contributing to
the $\et\pri$-propagator (not just the connected and the hairpin diagram as in
Q-QCD) and this propagator may even be well-defined in the field-theoretic
sense.

In summary, the fermions in ``topologically unquenched QCD'' are fully
dynamical, but they interact in a way which does not pay attention to the
details of the gluon background configuration but to its topological index
only and this seems to be sufficient to get a number of basic features
of full QCD qualitatively right.


\section*{Acknowledgements}
The author is supported by the Swiss National Science Foundation (SNF).


\end{document}